\documentclass[12pt,a4paper,dvips]{article}
\usepackage{a4p}
\usepackage{cite,mcite}
\usepackage{graphicx}
\usepackage{subfigure,rotating}

\def\as{\alpha_{\mathrm{s}}}
\def\etal{{\it et~al.}}
\def\Nf{\mathrm{N_{f}}}

\begin{document}
\begin{titlepage}
\begin{flushright}\vbox{\begin{tabular}{c}
           TIFR/EHEP-00/01  \\
	   IMSc-2000/06/21 \\
           June, 2000\\
           hep-ph/0006008\\
\end{tabular}}\end{flushright}
\begin{center}
   {\large \bf Quark Mass Corrections to the Perturbative Thrust and 
   its Effect on the determination of {\boldmath $\as$}}
\end{center}
\bigskip
\begin{center}
    Sunanda Banerjee\footnote{E-mail: sunand@tifr.res.in}\\
    {\em Tata Institute of Fundamental Research,\\
    Homi Bhabha Road, Mumbai 400005, India.}\\
    \medskip and\\ \medskip
    Rahul Basu\footnote{E-mail: rahul@imsc.ernet.in}\\
    {\em The Institute of Mathematical Sciences,\\
    Chennai (Madras) 600 113, India.}
\end{center}
\bigskip
\begin{abstract}
 We consider the effects of quark masses to the perturbative thrust 
 in $e^+e^-$ annihilation. In
 particular we show that perturbative power corrections resulting from
 non-zero quark masses considerably alters the size of
 the non-perturbative power corrections and consequently, significantly
 changes the fitted value of $\alpha_s$.
\end{abstract}
\end{titlepage}
One of the cleanest signatures of perturbative QCD comes from jet cross
sections in $e^+e^-$ annihilation. In such processes, it is possible to
define infra-red safe event shape variables which can be calculated
order by order in perturbative QCD and compared subsequently with
experiment. However in order to carry out these comparisons, a method has
to be evolved to parametrise non-perturbative effects which though 
expected to be small at present $Q^2$ values at LEP, actually turn out 
to be substantial ($\sim$ 25\%) even at $Q\sim m_{\mathrm{Z}}$. One of 
the reasons for this is that these 
non-perturbative effects are actually suppressed by a single power of
$Q$ rather than $Q^2$. In addition, it is also possible that these power
corrections could be comparable to ${\cal O}(\as^2)$ at present LEP
energies. 

In order to address these issues the Milan group of Dokshitzer \etal\
\cite{milan} drawing on the earlier work of Webber \cite{webber}, 
Korchemsky and Sterman \cite{ks} and others,  presented a systematic 
approach for
handling power corrections using perturbation theory. Very briefly, they
studied the consequences of assuming that $\alpha_s$ has a low energy
effective form which does not grow at low scales but has an infra-red regular 
form. The moments of $\as$ are integrated only over the infra-red region.
Various non-perturbative parameters are then parametrised and the form
and magnitude of power corrections are determined.

However before one uses the approach of the Milan group in order to get
a handle on power corrections and subsequently determine $\as$ by a
fit to the data, it is important to isolate power corrections coming
from a purely perturbative region. The Milan approach neglects the
masses of all the quarks but instead uses a gluon mass as a `trigger' to
differentiate the perturbative from the non-perturbative region. 
We find however that the masses of the quarks, particularly the {\bf c} 
and the {\bf b} quarks, even at present LEP energies, can contribute 
significantly (of the order of about 25\%). In fact, if we go beyond 
the top quark threshold (which is expected, perhaps in the future NLC) 
the perturbative contribution to power contributions due the top quark 
mass is even larger. We will have more to say on this later in the paper.

In this paper, we consider the example of one such event shape variable
- the thrust - and show the significance of the effect of quark masses
which need to be folded in before estimating the non-perturbative
contribution to power corrections. We present explicit expressions to
${\cal O}(\as)$ of quark mass corrections expanded to ${\cal O}(m)$. We 
also show the effect of keeping the full mass contribution to 
${\cal O}(\as)$
which unfortunately does not have a simple analytic form like the former
and needs to be calculated numerically. Using these expressions we then
fold in the power corrections of the Milan type and use this full
expression to estimate both $\alpha_0$ and $\as$ and compare it
with estimates that exist in the literature without taking quark masses
into account. 

The first paper which calculated the effect of quark masses to
${\cal O}(\as)$ was published about 16 years ago by one of the authors
\cite{rb}. For completeness, in what follows, we quote those results from
that paper which we need for our analysis here. The thrust, as defined
traditionally, is given by
\begin{equation}
T = 2\frac{\max\sum_{i\epsilon h}(p_i\cdot{\hat n})}{\sum_i\vert
p_i\vert},
\end{equation}
where the denominator runs over all observed particles and the numerator
runs over all particles in a hemisphere. ${\hat n}$ is a unit vector
chosen in a direction that maximises the numerator and defines the jet
axis. 

While this definition is appropriate for all massless particles, to
include mass effects in the definition of the thrust, we modify the
above definition slightly and write
\begin{equation}
T = 2 \frac{\max\sum_{i\epsilon h}(p_i\cdot {\hat n})}{W},
\end{equation}
where $W^2=s$. Of course the denominator equals $\sum_i \vert p_i\vert$
when all the particles are massless.  This normalisation with 
the total energy is also what is used by the Milan group
in their analysis though in their case the massive gluon eventually
decays into massless quarks and gluons. 

For a three particle final state, the thrust, as we define it, is given by
\begin{equation} 
T = \max\left[(x_1^2-\xi)^{1/2},(x_2^2-\xi)^{1/2},x_3\right],
\end{equation} 
where $x_i=2E_i/W$, $E_i$ being the energy of the $i$th particle in the
final state in the c.m. frame and $ \xi=4m^2/W^2$, $m$ being the mass of
the quarks. Note that in the two-jet limit $T=T_0\equiv\sqrt{1-\xi}$. 

The average value of the thrust is defined by
\begin{equation} 
<T> = \frac{\left[\int T\frac{d\sigma}{dT}dT\right]}{\left[\int
\frac{d\sigma}{dT}dT\right]}.
\end{equation} 
The numerator of the above is given up to ${\cal O}(\as)$ and to 
${\cal O}(\xi)$ 
by ($\sigma_0 = (4\pi\alpha_2/s)e_i^2$ is the total cross section 
for $e^+e^- \rightarrow q_i\bar q_i$) \cite{rb}
\begin{eqnarray} 
\frac{1}{\sigma_0}\int T\frac{d\sigma}{dT}&=&1-\frac{\xi}{2}+
\frac{4\as}{3\pi}\left\{\frac{137}{16}\xi+\frac{5}{4}\xi\ln 2
-\frac{1}{2}\xi^{1/2}+\frac{7}{9}+\frac{1}{4}\xi\ln^2 \xi - \xi\ln 2\ln \xi
+ \frac{\pi^2}{6}\right. \nonumber \\
&&-\frac{\xi\pi^2}{6} - \frac{1}{8}\xi\ln\xi 
-\frac{1}{2}\xi\ln 3\ln 2 - \frac{9}{2}\xi\ln 3-\ln^2 3+\frac{3}{8}\ln 3
\nonumber \\
&&-\frac{1}{3}\xi\left[Li_2(1-\xi^{1/2}+\xi/2) 
-Li_2(\frac{1}{3}+\frac{1}{2}\xi) \right] 
 -2Li_2(\frac{2}{3}-\frac{1}{2}\xi) \nonumber \\
&& \left. +\xi Li_2(\frac{2}{3}-\frac{1}{2}\xi)
- \frac{1}{2}\xi Li_2(\frac{1}{3}-\frac{1}{4}\xi)+\xi\ln^2 2\right\},
\end{eqnarray}
where $Li_2(x)$ is the dilogarithm function. In the $\xi\rightarrow 0$
limit this gives, for the average thrust,
\begin{equation}
<T> = 1+\frac{4\as}{3\pi}\left[\frac{1}{36}+\frac{\pi^2}{6}-\ln^2 3
+\frac{3}{8}\ln 3 - 2 Li_2(\frac{2}{3})\right],
\end{equation}
which works out to, for the perturbative thrust in the massless limit,
\begin{equation}
<1-T>  =  1.05\frac{\as}{\pi},
\end{equation}
as quoted in numerous places in the literature. 

Several points here are worthy of note. The leading term in the $O(m)$
expansion above is $\xi^{1/2}$. Thus the leading mass correction goes as
$1/Q$. To the best of our knowledge, this fact was noticed for the first
time in \cite{rb} and subsequently in \cite{webber} and \cite{milan} who
have traced it to appear from the soft phase space boundary. We would
like to stress that this $1/Q$ behaviour is a pure perturbative higher
twist effect to the thrust and not related to any non-perturbative 
contribution. Thus, it seems clear, that the coefficient of $1/Q$ in the 
full expression for the thrust would include contributions both from the
perturbative as well as the non-perturbative sectors. This aspect will
become more quantitative, when we do our fits later.

The second point to note is a calculational one. Since, in the two jet
limit, the thrust is equal to $T_0=\sqrt{1-\xi}$, in order to make the
virtual contributions vanish we need to calculate, not as in the usual
case $<1-T>$, but $<T_0 - T>$. It is then a trivial matter to add a term
$<1-T_0>$ to obtain $<1-T>$ to compare with experiment. 

In order to compare with experiment, however, and to redo the fits for
$\as$ and $\alpha_0$ we have used not only the ${\cal O}(m)$ contribution
above but also the full massive contribution to ${\cal O}(\as)$ albeit
evaluated numerically. In addition we have also compared the
${\cal O}(\as^2)$ massless corrections to the ${\cal O}(\as)$ massive
correction to try and estimate how much of the $1/Q$ corrections can be
mimicked by higher orders in the coupling constant. 

\begin{figure}[htb]
\begin{center}
 \includegraphics[width=0.9\linewidth]{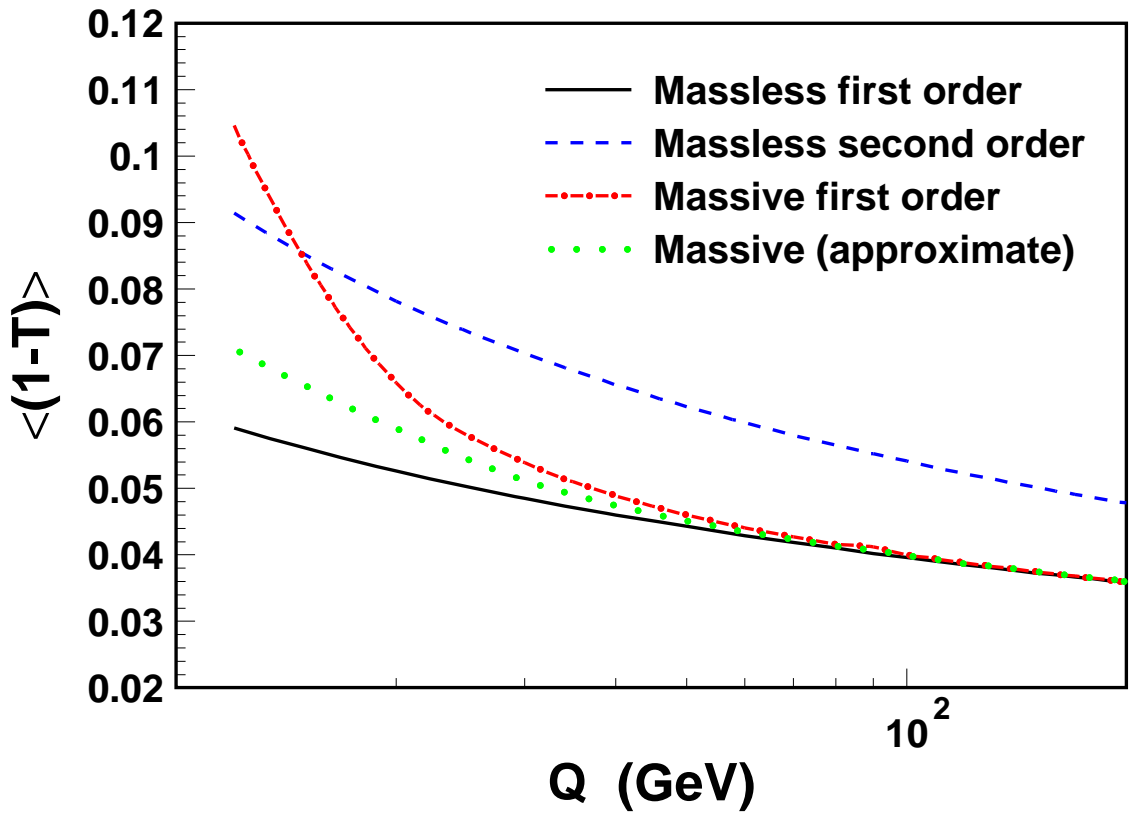}
\end{center}
\caption[]{Effect of mass correction and higher order in the mean thrust
           value. The solid and the dashed lines are respectively the 
           first and second order calculations of $<1-T>$ without mass 
           corrections. The dashed-dotted and dotted lines are complete
           and approximate (${\cal O}(m)$) first order calculations with 
           quark mass effects.}
\label{fig:thrust}
\end{figure}

Figure \ref{fig:thrust} shows $<1-T>$ as a function of centre of mass
energy computed with $\as(m_{\mathrm{Z}})$ = 0.12. As one sees from 
the curve the contribution of the second order terms are large over
the entire energy region ($\sim$ 55\% at $Q$ = 12 GeV going down to 33\% 
at 200 GeV). On the other hand the effect of quark masses, evaluated only
to first order in $\as$, is even larger at low centre of mass energy ($\sim$ 
76\% at $Q$ = 12 GeV). This is clearly a $1/Q$ power law effect and 
hence dies off faster, becoming 2.5\% at 
$Q \sim m_{\mathrm{Z}}$ and negligible at 200 GeV. It is clear from the
figure that one needs to take the full massive correction rather than
the ${\cal O}(m)$ contribution, because it accounts only 60\% (30\%) of 
the mass correction at 20 GeV (12 GeV).

In order to compare the theoretical predictions with the measurements
done at different centre of mass energies \cite{data} at PETRA, PEP, 
TRISTAN, SLC and LEP, we add the non-perturbative contribution {\em a
la} the Milan group \cite{milan} to the perturbative contribution. In this
paper, we use only the ${\cal O}(\as)$ calculation of $<1-T>$ and 
a more detailed comparison with a ${\cal O}(\as^{2})$ calculation is
under preparation \cite{rbp}. We will have more to say on this later.
The non-perturbative contribution, as is well known, is
given by an additive contribution $<1-T>_{\mathrm{pow}}$:
\begin{equation}
<1-T>_{\mathrm{pow}} = 2 \frac{4C_{F}}{\pi^{2}} {\cal{M}} 
 \frac{\mu_{I}}{Q} \left[ \alpha_{0}(\mu_{I}) - \as(Q) - \beta_{0}
 \frac{\as^{2}(Q)}{2\pi} \left( \ln \frac{Q}{\mu_{I}} + 
 \frac{K}{\beta_{0}} + 1\right)\right]
\end{equation}
where $\mu_{I}$ is an infra-red matching scale (taken as 2 GeV), $K$ = 
(67/18 $-$ $\pi^{2}$/6)$\cdot C_{A}$ $-$ $5\Nf/9$ and $\cal{M}$ is the
Milan factor (determined to be 1.49) \cite{milanfactor}.

\begin{figure}[htb]
\begin{center}
 \includegraphics[width=0.9\linewidth]{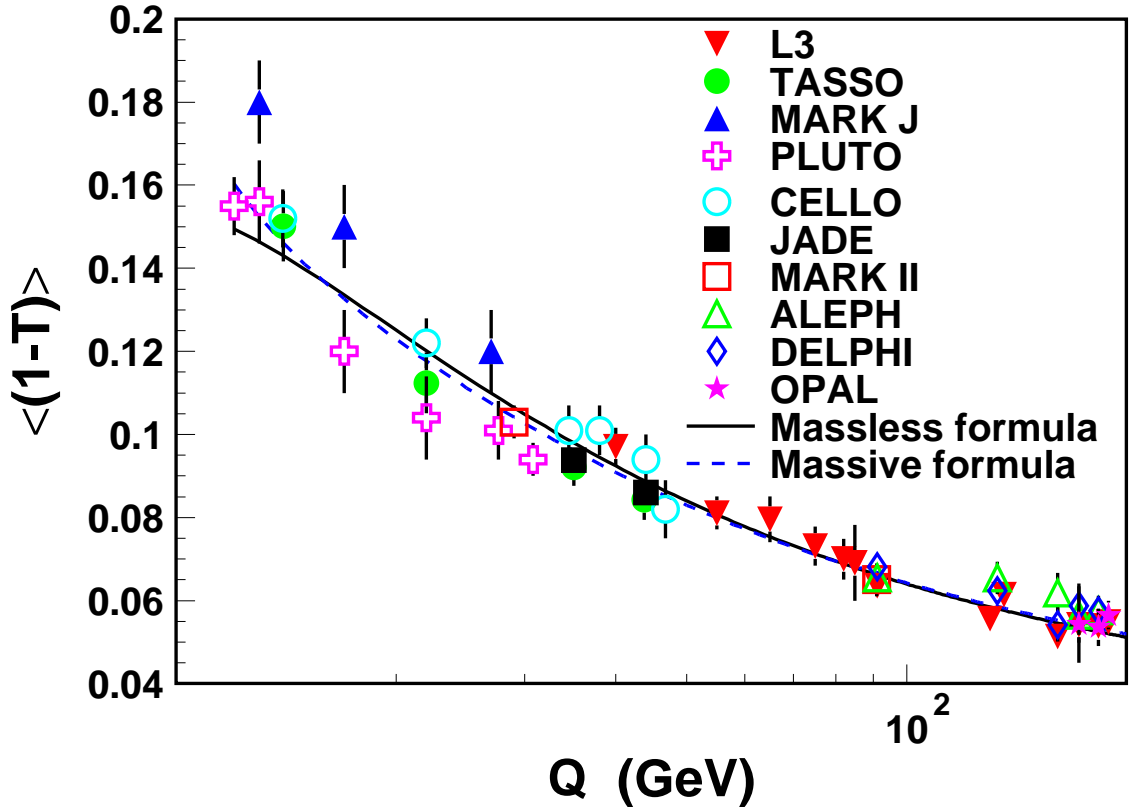}
\end{center}
\caption[]{Results of the fit of $<1-T>$ to the first order calculations 
           with and without quark mass effects. The points indicate 
           measured $<1-T>$ values from different experiments and the
           solid and dashed lines are fits without and with mass 
           corrections.}
\label{fig:thrustfit}
\end{figure}

Figure \ref{fig:thrustfit} shows the experimental values of $<1-T>$
together with the two fits which use respectively the massless and 
the massive forms (both to ${\cal O}(\as)$ for the perturbative 
contribution). These fits have
been carried out with two free parameters $\as (m_{\mathrm{Z}})$ and
$\alpha_{0}$. Both massless and massive formulation of the perturbative
component give reasonable fits to the data with $\chi^{2}$ of 90.2 and 
69.2 respectively for 48 degrees of freedom corresponding to confidence
levels of $0.22\times 10^{-3}$ and $0.24\times 10^{-1}$. However, they 
do differ in the final values of $\as (m_{\mathrm{Z}})$ and $\alpha_{0}$
as can be seen in Table \ref{tab:fit}. In these fits the scale parameter
is chosen to be 1.0.

\begin{table}[htb]
\begin{center}
\begin{tabular}{|l|c|c|}\hline
 Fit Type & $\alpha_{0}$ & $\as (m_{\mathrm{Z}})$ \\ \hline
Massless quarks & $0.7931 \pm 0.0066$ & $0.1533 \pm 0.0015$ \\
Massive  quarks & $0.7385 \pm 0.0065$ & $0.1612 \pm 0.0014$ \\ \hline
\end{tabular}\end{center}
\caption[]{Results of the fits of the $<1-T>$ distribution to
           perturbative and power law terms when the quark mass
           effects in the perturbative term is ignored or included}
\label{tab:fit}
\end{table}

The errors quoted in the Table \ref{tab:fit} are experimental errors
obtained from the minimisation procedure. We can also estimate the theoretical 
uncertainties on these quantities by varying the scale parameter.
If we vary the scale parameter between 0.5 and 2.0, we obtain uncertainties 
in $\as$ and $\alpha_{0}$ to be $\pm 0.010$ and $\pm 0.12$ respectively.
The value of $\as (m_{\mathrm{Z}})$, obtained from the fits, when
quark mass effects are included or ignored, differ by 0.008 which is
much larger than the experimental uncertainty of about .001 on the $\as$ 
value and comparable in fact to the theoretical uncertainty.

It is thus clear from the preceding analysis that an estimate of the
power corrections due to the non-zero masses of the quarks is crucial in
getting better and more realistic estimates on the strong coupling
constant and indeed, in general, on power corrections. The next obvious
step would be to calculate mass corrections to ${\cal O}(\as^2)$. Some
results in this direction have been obtained by Nason and Oleari
\cite{no} which could be used to carry out a similar analysis to the one
presented above. We are, at present, in the process of extending our
analysis to second order in the strong coupling using the results of
\cite{no}.

In various projected Linear Collider scenarios (like, for example, the
NLC) energies upwards of 500 GeV are expected. In such a region, the
effect of the top quark would be dramatic and significant. The
combination of the large mass of the top quark and a charge squared of
4/9 implies that the usual massless expressions for the thrust would not
work. We have estimated that the difference between choosing a massless
formula for describing the thrust beyond the top quark threshold and
using the (more appropriate) massive formula changes the value of the
thrust by about a factor of 5 near the threshold. Most of this contribution 
comes, in fact, from the top quark mass.
In the table below we give an estimate of the change that would occur
between choosing all quarks massless and massive above the top quark
threshold. It is obvious that the effect is spectacularly large,
particularly near the threshold. Mass effects in the resummation of
event shape variables are also expected to be significant and this is
presently being studied. 

\begin{table}[htb]
\begin{center}
\begin{tabular}{|c|c|c|c|}\hline
 Q (GeV) & $\alpha_{s}$ & $<1-T>$(massless)&$<1-T>$(Massive) \\ \hline
360 & .0995 & .0323 & .1605 \\
500 & .0957 & .0311 & .0994 \\
1000 & .0888 & .0289 & .0435 \\ \hline
\end{tabular}\end{center}
\caption[]{Difference between choosing massless and massive quarks above
the top quark threshold}
\end{table}

Thus, it is imperative that
in order that reliable estimates be made of the thrust at these
energies, we have available, calculations to higher orders in $\as$
of $e^+e^-$ scattering with massive quarks in the final state. This
would also give us a handle on the relative magnitudes of power
corrections to the thrust to a particular order in $\as$ and the 
magnitude of the next order term in $\as$ \cite{milanfactor, sterman}. 
For example NNLO effects might be capable of mimicking the $1/Q$
behavior. Mass effects in the resummation of
event shape variables are also expected to be significant and this is
presently being studied. 

\bigskip
\noindent {\bf Acknowledgements}: One of us (R.B.) would like to thank George
Sterman for many illuminating discussions. This project was started at
the Sixth Workshop on High Energy Physics Phenomenology (WHEPP-6),
held at Chennai, India in January 2000 and we would like to thank all the
funding agencies which made this workshop possible.


\bigskip \bigskip
\begin{thebibliography}{99}
\bibitem{milan}
Yu. L. Dokshitzer and B. R. Webber, Phys. Lett. {\bf B352} (1995) 451;\\
Yu. L. Dokshitzer, G. Marchesini and B. R. Webber, Nucl. Phys. {\bf B469}
(1996) 93.
\bibitem{webber}
B. R. Webber, Phys. Lett. {\bf B339} (1994) 148. 
\bibitem{ks}
G. P. Korchemsky and G. Sterman, Nucl. Phys. {\bf B437} (1995) 415.
\bibitem{rb}
R. Basu, Phys. Rev. {\bf D29} (1984) 2642.
\bibitem{data}
 ALEPH Collaboration, D. Bukusulic $\etal$, Z. Physik {\bf C55} (1992) 
 209. \\
 ALEPH Collaboration, D. Bukusulic $\etal$, ALEPH 98-025 (1998),
 contributed to 1998 winter conferences. \\
 AMY Collaboration, Y. K. Li $\etal$, Phys. Rev. {\bf D41} (1990) 2675.\\
 CELLO Collaboration, H. J. Behrend $\etal$, Z. Physik {\bf C44}
 (1989) 63. \\
 DELPHI Collaboration, P. Abreu $\etal$, Phys. Lett. {\bf B456}
 (1999) 322. \\
 JADE Collaboration, A. Morivilla Fernandez $\etal$, Eur. Phys.
 J. {\bf C1} (1998) 1. \\
 L3 Collaboration, M. Acciari $\etal$, CERN-EP/2000-064 (2000). \\
 MARK-II Collaboration, A. Peterson $\etal$, Phys. Rev. {\bf D37}
 (1988) 1. \\
 MARK-II Collaboration, A. Abrams $\etal$, Phys. Rev. Lett. {\bf 63}
 (1989) 1558. \\
 MARK-J Collaboration, D. P. Barber $\etal$, Phys. Rev. Lett. 
 {\bf 43} (1979) 902. \\
 OPAL Collaboration, G. Abbiendi $\etal$, CERN-EP/1999-178 (1999).\\
 PLUTO Collaboration, Ch. Berger $\etal$, Z. Physik {\bf C12}
 (1982) 297. \\
 TASSO Collaboration, W. Braunschweig $\etal$, Z. Phys. {\bf C47}
 (1990) 187.
\bibitem{rbp}
 Under preparation.
\bibitem{milanfactor}
  Yu. L. Dokshitzer, hep-ph/9911299.
\bibitem{no}
P. Nason and C. Oleari, Nucl. Phys. {\bf B521} (1998) 237.  
\bibitem{sterman}
G. Sterman, talk at the QCD-Euronet workshop, Florence, Italy, October
1999.
\end{thebibliography}
\end{document}